\documentclass[prd,showpacs,superscriptaddress,nofootinbib,floatfix,11pt]{revtex4-2}
\usepackage[utf8]{inputenc}
\usepackage{amsmath,amssymb}
\usepackage{slashed}
\usepackage{subfigure}
\usepackage[colorlinks=true,
breaklinks=true,
urlcolor=magenta,
citecolor=blue]{hyperref}
\usepackage[usenames,dvipsnames]{color}
\usepackage{appendix}
\usepackage{braket,bm}
\usepackage{multirow}
\usepackage{enumitem}
\usepackage{color}
\usepackage{slashed}
\usepackage{orcidlink}
\usepackage{graphicx}
\usepackage{tikz}
\usepackage{booktabs}
\usepackage{array}
\usepackage{overpic}
\usepackage{float}
\usepackage{threeparttable}
\allowdisplaybreaks[4]
\newcommand{\zjwl}{\affiliation{College of Information and Intelligence Engineering, Zhejiang Wanli University, Zhejiang 315101, China}}

\newcommand{\nbu}{\affiliation{Physics Department, Ningbo University, Zhejiang 315211, China}}

\newcommand{\usc}{\affiliation{School of Nuclear Science and Technology, University of South China, Hengyang, 421001, Hunan, China}}

\newcommand{\kmc}{\affiliation{School of Physics Science and Technology, Kunming University, Kunming 650214, China}}

\begin{document}
\title{Contributions of interference and non-interference components to CP asymmetries in heavy meson decays} 
\author{Jing-Juan Qi\orcidlink{0000-0002-9260-9408}}\email{jjqi@mail.bnu.edu.cn}
\zjwl\nbu
\author{Yi-Fan Zhao}
\nbu
\author{Jin-Xia Liu}
\zjwl

\author{Zhen-Hua Zhang\orcidlink{0000-0001-5031-9499}}\email{zhangzh@usc.edu.cn}
\usc

\author{Zhen-Yang Wang\orcidlink{0000-0002-4074-7892}}\email{Corresponding author: wangzhenyang@nbu.edu.cn}
\nbu

\author{Xin-Heng Guo\orcidlink{0000-0002-9309-9112}}\email{Corresponding author:xhguo@bnu.edu.cn}
\kmc

\begin{abstract}	
In multi-body decays of heavy mesons, conventional CP asymmetry observables obtained by integrating over the full phase space are insensitive to the higher-order wave expansion  contributions in the decay amplitude squared, and consequently fail to retain information on interference effects among different resonances. To overcome this  limitation, one can introduce a phase-space partitioning scheme based on the zeros of Legendre polynomials, supplemented by a sign-function weighting procedure. On such a basis,  two observables are defined, namely an asymmetry observable 
$\mathcal{A}_{\pm}^{\mathrm{asy},l}$,  and the corresponding CP asymmetry 
$\mathcal{A}_{\mathrm{CP}}^{\mathrm{asy},l}$. We further separate the observables into interference and non-interference parts and analyze their respective roles. As an application,  the decay channel 
$B^\pm\rightarrow\pi^\pm\pi^+\pi^-$  are analyzed in the region near the $\rho^0(1450)$ resonance. Using the LHCb data, the results show that odd-$l$
 schemes are particularly effective in isolating interference contributions, while even-$l$ schemes are more sensitive to non-interference terms. This new assignment scheme has the potential to be extended to other decay processes, thus enriching the available physical observables.

\end{abstract}
	
	\date{\today}	
	\maketitle
\newpage

\section{Introduction}\label{sec1}
CP violation (CPV) was first discovered in 1964 through studies of  the neutral kaon system \cite{Christenson:1964fg}. It is an important feature of the weak interaction and is related to the matter–antimatter asymmetry in the Universe \cite{Kobayashi:1973fv}. In the
Standard Model (SM), CPV results from the
weak complex phase in the Cabibbo-Kobayashi-Maskawa
(CKM) matrix that reflects the transitions of different
generations of quarks.  CPV in decay processes requires at least two interfering amplitudes with distinct strong and weak phases. The weak phases, related to the complex elements of the CKM matrix, change  the sign under the CP conjugation,  whereas the strong phases, which may arise  from hadronic final-state interactions, remain  invariant. Experimental discovery of CPV has been established in many decays \cite{Christenson:1964fg,KTeV:1999kad,BaBar:2001pki,Belle:2001zzw,Belle:2010xyn,BaBar:2010hvw,LHCb:2013syl,LHCb:2019hro,LHCb:2025ray}. 
The direct CPV is the most widely studied form of CPV. Some interference terms vanish after the differential decay widths are integrated over the whole phase space, yet they can appear in angular observables \cite{Donoghue:1987wu,Valencia:1988it,Kayser:1989vw,Bensalem:2000hq,Wang:2022fih}. For a $B$ or $D$ three-body decay, within the partial wave analysis, 
the general expression for such a CP asymmetry (CPA) observable is
\begin{equation} 
\label{ACPP1}
\mathcal{A}_{CP}\propto \sum_{l=0}^n\int w_-^l(s_{12})ds_{12}P_l(\cos\theta)d\cos\theta-\sum_{l=0}^n\int w_+^l(s_{12})ds_{12}P_l(\cos\theta)d\cos\theta,\\
\end{equation}
where $w_\pm^l(s_{12})$ denote  the weights in the Legendre expansion of the decay amplitude squared
for a pair of CP-conjugate processes, $P_l(\cos\theta)$ is the $l$-th Legendre polynomial, $\theta$ is the angle between two same-sign pions or kaons in the rest frame of the intermediate resonance. 

In the conventional treatment, once the  angular variable $\cos\theta$ is integrated over the unbroken interval $[-1,1]$, all terms $w_{\pm}^l$ with $l\geq1$ vanish because of the orthogonality of the Legendre polynomials, and 
only the zeroth-order  terms $w_+^0$ and $w_-^0$  survive (see Eqs. (\ref{Mtott4}) and (\ref{Mtott21})). As a result, the contributions from higher-order wave expansion terms
are completely lost. This implies, in particular, that the interference effects among resonances with different quantum numbers cannot be fully explored in such a treatment. To overcome this limitation,  we adopt a phase-space partitioning approach that subdivides the angular integration domain according to the zeros of Legendre polynomials, with alternating signs assigned to adjacent subintervals \cite{Zhang:2022emj,Hu:2022eql,Zhang:2021fdd,Qi:2024zau,Qi:2025zna}. Two novel physical quantities are introduced, the asymmetry and the corresponding CP asymmetry observables. Then we can further investigate the contributions from both interference and non-interference components, respectively.

The LHCb collaboration has reported  detail  amplitude analyses of  the $B^\pm\rightarrow\pi^\pm\pi^+\pi^-$ decays  \cite{LHCb:2019sus,LHCb:2019jta}. In the low $\pi^+\pi^-$ invariant mass region,  CPV  was  established through the interference between the $\pi^+\pi^-$ $S$-wave and $P$-wave amplitudes, which was further supported by \cite{Wei:2022zuf,Cheng:2022ysn}. In the higher $\pi^+\pi^-$ invariant mass region, significant CPAs were also observed in the $\pi^+\pi^-$ $S$- and $D$-wave components, as well as in the interference between  $S$- and $P$-wave resonances.  The collaboration reported the first observation of a CPA involving the $f_2(1270)$ resonance, consistent with the theoretical predictions \cite{Cheng:2010yd,Chang:2024qxl,Li:2018lbd,Zou:2012td}. Related studies had also been reported earlier by the $BABAR$ collaboration {\cite{BaBar:2005jqu,BaBar:2009vfr}. However, the pronounced $S$-wave behavior observed around 1.5 $\mathrm{GeV/c}^2$ cannot be definitively linked to the $f_0(1500)$ resonance as expected in the $K$-matrix model \cite{LHCb:2019sus}.    Compared with the low resonance mass region, theoretical studies of the interference mechanisms between different resonance states in the higher mass region remain relatively limited, possibly because of the complex structure of the higher excited states. In this work, we will focus on  the vicinity of the $\rho(1450)^0$ resonance, where overlap with other  resonances could be important. By testing different hypotheses, 
we aim to assess the possible role of the
$f_0(1500)$ and to study the two observables introduced above.

This paper is organized as follows. In Sec. ${\mathrm{\uppercase\expandafter{\romannumeral2}}}$, we will introduce new observables which include  higher-order wave expansion terms, and the CP asymmetry induced by these observables.
In Sec. ${\mathrm{\uppercase\expandafter{\romannumeral3}}}$,  we will apply the framework to $B^\pm\rightarrow\pi^\pm\pi^+\pi^-$, with the two subsections presenting the detailed theoretical analysis and the numerical results, respectively. We will briefly give the summary in Sec. IV.

\section{The higher-order wave expansion term induced asymmetry, and its induced CP asymmetry}\label{Definitions}

The cascade decay approach is typically employed to deal with the multi-body decay $M\rightarrow M_1M_2M_3 \cdots$, where $M\rightarrow R_1M_3 \cdots \rightarrow M_1M_2M_3 \cdots$. When we focus on the mass region near $R_1$, the contributions from its neighboring resonances $R_2$, $R_3$, $\cdots$ should also be taken into account.  For three-body decays, within the partial wave framework, when the resonances corresponding to varying orbital angular momenta $l$ are taken into account, the total amplitude can be constructed in terms of the Legendre polynomials $P_{l}(\cos\theta)$ as

\begin{equation} \label{Mtot2}
\mathcal{M}_{\pm}^{\mathrm{tot}}(s_{12},\cos\theta)=\sum_{l}
\mathcal{M}_{\pm}^{l}(s_{12})P_{l}(\cos\theta),\\
\end{equation}
where the subscripts  $``\pm"$ represent the total decay amplitudes for the two CP-conjugate processes of the parent particle $M$, $p_i (i=1,2,\cdots)$ are the 4-momenta of the  particales in the final states, $s_{12}=(p_1+p_2)^2$, $\cos\theta=\frac{\vec{p}_1\cdot \vec{p}_3}{|\vec{p}_1||\vec{p}_3|}=\frac{s_{23}-(s_{23,\mathrm{max}}+s_{23,\mathrm{min}})/2}{(s_{23,\mathrm{max}}-s_{23,\mathrm{min}})/2}$ with $s_{23}=(p_2+p_3)^2$ and  $s_{23,\mathrm{min}(\mathrm{max})}$ being the minimum (maximum) values of $s_{23}$, respectively, for fixed $s_{12}$. 

The  total amplitude squared can be  expressed in the form of Legendre polynomials, as shown:
\begin{equation} \label{Mtotsqu1}
\begin{split}
|\mathcal{M}_{\pm}^{\mathrm{tot}}(s_{12},\cos\theta)|^2=\sum_{l}w_\pm^l(s_{12})P_l(\cos\theta),\\
\end{split}
\end{equation}
where $w_\pm^l(s_{12})$ 
are the corresponding weights in the expansion and can be extracted by using the orthogonality of Legendre polynomials, which can be extracted using the 
$w_{\pm}^l(s_{12})=\int_{-1}^1|\mathcal{M}_{\pm}^{\mathrm{tot}}(s_{12},\cos\theta)|^2P_l(\cos\theta)d\cos\theta$.

The  orthogonality property of Legendre polynomials, 
\begin{equation}
\label{eq:op}
\int_{-1}^1P_l(\cos\theta)P_{l'}(\cos\theta)d\cos\theta=\frac{2}{2l+1}\delta_{ll'},
\end{equation} 
ensures that different partial waves can be separated. However, when integrating over the full angular range $[-1,1]$
, all terms with 
$l\geq1$ vanish due to
\begin{equation}
\label{eq:op}
\int_{-1}^1P_l(\cos\theta)d\cos\theta=0 \quad \text{for} \quad l\geq1.
\end{equation} 

Consequently, conventional CPA observables are insensitive to interference effects between different partial waves. To recover this lost informationand probe the corresponding observable sensitivity, we subdivide the angular integration range using the zeros of Legendre polynomials. For the $l$-th order Legendre polynomial $P_l(\cos\theta)$, there are $l$ zeros in the interval $[-1,1]$. We denote these zeros as $x_1^{(l)}<x_2^{(l)}<\cdots<x_l^{(l)}$, which divide the interval into $l+1$
subintervals: 
\begin{equation}
I_0=[-1,x_1^{(l)}],\quad I_i=[x_i^{(l)}, x_{i+1}^{(l)}] \quad \text{for} \quad i=1,\cdots,l-1, \quad I_l=[x_l^{(l)},1].
\end{equation} 

Within each subinterval, $P_l(\cos\theta)$ maintains a constant sign, while adjacentsubintervals exhibit alternating signs. To quantify the effects of these behaviors, one can define an asymmetry observables as follows:

\begin{equation}
\label{eq:hatACPl}
\mathcal{A}_{\pm}^{\mathrm{asy},l}= \frac{\sum_{i=0}^l(-1)^{i+1}\int_{I_i} \left|\mathcal{M}^{\mathrm{tot}}_{\pm}(s_{12},\cos\theta)\right|^2\tilde{d}\cos\theta ds_{12}}{\sum_{i=0}^l\int_{I_i} \left|\mathcal{M}_{\pm}^{\mathrm{tot}}(s_{12},\cos\theta)\right|^2\tilde{d}\cos\theta ds_{12} },
\end{equation} 
where  $\tilde{d}\cos\theta=(s_{23,\mathrm{max}}-s_{23,\mathrm{min}})/2d\cos\theta$. A more intuitive description of the sign convention can be achievedby introducing the sign function sgn(z) ($=\pm1$ when $z\gtrless0$), and  we can get

\begin{equation}
\label{eq:hatACPl}
\mathcal{A}_{\pm}^{\mathrm{asy},l}=\frac{\sum_j(\mathcal{N}_{\pm}^j)\text{sgn}(P_l(\cos\theta_j))}{\mathcal{N}_{\pm}},
\end{equation} 
where $\mathcal{N}_{\pm}^j$ are  the event yields of the CP-conjugate processes in each bin $j$, which are  
\begin{equation}
\label{events}
\mathcal{N}_{\pm}^j=R\int\left|\mathcal{M}_{\pm}^{\mathrm{tot}}(s_{12},\cos\theta)\right|^2_{s_{12}=s_{12,i}}\tilde{d}\cos\theta,    
\end{equation}
where $R$ is the phase-space factor, and $s_{12,j}$ is the central value of $s_{12}$ of bin $j$ .

To describe the distinction of asymmetry between the pair of CP-conjugate processes, we define the CPA as the following:
\begin{equation}
\label{AasyCP}
\mathcal{A}_{\mathrm{CP}}^{\mathrm{asy},l}=\frac{1}{2} (\mathcal{A}_{-}^{\mathrm{asy},l}-\mathcal{A}_+^{\mathrm{asy},l}).   
\end{equation}

{
\begin{table}[tb]
\scriptsize
\renewcommand{\arraystretch}{1.0}
\centering
\caption{The  classifications, masses (in MeV), and decay widths (in MeV) of the resonances \cite{ParticleDataGroup:2024cfk}.}
\begin{tabular*}{\textwidth}{@{\extracolsep{\fill}}ccc|ccc|ccc}
\hline
\hline
Scalar   & Mass       & Decay width   &Vector   &Mass    & Decay width    &Tensor     & Mass &  Decay width  \\
\hline
 $f_0(1370)$  &$1200\sim1500$    &$200\sim500$&$\rho^0(1450)$&$1465\pm25$&$400\pm60$ &$f_2(1270)$         &$1275.4\pm0.8$    &$185.8^{+2.8}_{-2.1}$\\
$f_0(1500)$ & $1522\pm25$  &$108\pm33$ &&&  &$f_2(1430)$  &$\approx1430$ &$46\pm15$ \\
 &    &&&&   &$f_2'(1525)$   &$1517.3\pm2.4$ &$84.4\pm2.7$\\
 &  &&&& &$f_2(1565)$    &$1571\pm13$&$132\pm23$\\
\hline
\hline
\end{tabular*}\label{Pl}
\end{table}}

\section{Application to $B^\pm\rightarrow\pi^\pm\pi^+\pi^-$ decay}\label{B to 3pi}
\subsection{Theoretical Analysis}\label{Theoretical Analysis}
In this section, we apply the theoretical framework introduced above to the $B^\pm\rightarrow\pi^\pm\pi^+\pi^-$ decay, focusing on the region near the $\rho^0(1450)$ resonance, which width is relatively large i.e., about 400 MeV. This broad width necessitates a careful treatment of overlapping contributions from nearby resonances. We will restrict the analysis to an observation window of one-half width around $\rho^0(1450)$, i.e., $s_{12}=[(m_{\rho^0}-\Gamma_{\rho^0}/2)^2, (m_{\rho^0}+\Gamma_{\rho^0}/2)^2]$, and will systematically examine possible contributions from $f_2(1270)$, $f_0(1370)$, $f_2(1430)$, $f_0(1500)$, $f'_2(1525)$ and $f_2(1565)$ which appear in this window. Their classifications, masses, and decay widths  are summarized in Table \ref{Pl}. 
The LHCb collaboration observed a significant $D$-wave signal, which was identified as the $f_2(1270)$ resonance. For the  $f_2(1430)$ ($\Gamma_{f_2(1430)}=46$ MeV) and $f'_2(1525)$ ($\Gamma_{f'_2(1525)}=72$ MeV) resonances, both of them are too high in mass and too narrow to be likely
to induce a significant effect in the region of interest.  This observation is further supported for $f'_2(1525)$ by its strongly suppressed 
$\pi\pi$ branching fraction, $\mathcal{B}(f'_2(1525)\rightarrow \pi\pi)\approx 8\text{\textperthousand}$, compared with  $\mathcal{B}(f'_2(1525)\rightarrow KK)\approx89\%$ and $\mathcal{B}(f'_2(1525)\rightarrow\eta\eta)\approx10\%$. In addition, dedicated fits are performed in which $f_2(1270)$ is replaced by either $f_2(1430)$ or  $f'_2(1525)$ within the two schemes given below, confirming that the contributions from $f_2(1430)$ and   $f'_2(1525)$ can be neglected. For similar reasons, the tensor resonance $f_2(1565)$ is also ignored, 
 although its width $\Gamma_{f_2(1565)}=130$ MeV is comparable to $\Gamma_{f_2(1270)}=186.6$ MeV. To assess possible scalar contributions, we consider two representative schemes.  Scheme 1 (S1) excludes the contributions from $\pi\pi$ S-wave components and only includes the interference between $P$- and $D$-wave resonances, 
 $\rho^0(1450) (l=1)$ and $f_2(1270) (l=2)$.
 Scheme 2 (S2) additionally incorporates the contribution from
$f_0(1370) $  or $f_0(1500) (l=0)$ into S1, to  determine whether a scalar contribution is required and which one provides a more appropriate description.
{\begin{table}
\renewcommand{\arraystretch}{1.2}
  \begin{center}
    \caption{The projection weights of every interference and non-interference component for the $w_{\pm}^l$.}
\begin{tabular*}
{\textwidth}{@{\extracolsep{\fill}}c|cccccc}
    \hline\hline
    \multirow{2}{*}{$l$} &\multicolumn{3}{c}{Non-interference terms } & \multicolumn{3}{c}{Interference terms }  \\
    \cline{2-4}
   \cline{5-7}
&$|\mathcal{M}_{\pm}^{f_2}|^2$ & $|\mathcal{M}_{\pm}^{\rho_0}|^2$  & $|\mathcal{M}_{\pm}^{f_0}|^2$ & $\mathrm{Re}[\mathcal{M}_{\pm}^{\rho_0}{\mathcal{M}_{\pm}^{f_0}}^*]$ & $\mathrm{Re}[\mathcal{M}_{\pm}^{\rho_0}{\mathcal{M}_{\pm}^{f_2}}^*]$ &  $\mathrm{Re}[\mathcal{M}_{\pm}^{f_0}{\mathcal{M}_{\pm}^{f_2}}^*]$  \\
    \hline
    {$l=0$} &  $1/5$   &  $1/3$ & $1$  & $-$ &$-$  &$-$\\
        \hline
    {$l=1$} &  $-$      & $-$  &  $-$    & $2$   &  $4/5$  &$-$\\
        \hline
{$l=2$} &  $2/7$     &  $2/3$  & $-$    & $-$   &  $-$& $2$ \\
        \hline
    {$l=3$} &  $-$     & $-$  & $-$    & $-$   & $6/5$   &$-$\\
        \hline
    {$l=4$} &  $18/35$      & $-$   & $-$    & $-$   & $-$ & $-$ \\
    
 \hline\hline
\end{tabular*}\label{tabw}
 \end{center}
  \end{table}}

The analysis is initially performed within the framework  S2, because  S1 represents a special case of S2 with $\mathcal{M}_\pm^{f_0}=0$.  In S2, the total amplitude can be expressed as
\begin{equation} \label{Mtott1}
\begin{split}
\mathcal{M}_{\pm}^{\mathrm{tot}}=\mathcal{M}_{\pm}^{f_0}P_0(\cos\theta)+\mathcal{M}_{\pm}^{\rho^0}P_1(\cos\theta)+\mathcal{M}_{\pm}^{f_2}P_2(\cos\theta),\\
\end{split}
\end{equation}
where $f_0$ represents either the $f_0(1370)$ or $f_0(1500)$ resonance, $\rho^0$ and $f_2$ are used as the abbreviated forms for  $\rho^0(1450)$ and $f_2(1270)$, respectively. We can abbreviate $\mathcal{M}_{\pm}^{f_0}$, $\mathcal{M}_{\pm}^{\rho^0}$  and $\mathcal{M}_{\pm}^{f_2}$ as $\mathcal{M}_{\pm}^{R}$, which can be factorized as: 
\begin{equation}\label{MRj}
\begin{split}
\mathcal{M}_{\pm}^{R}(s_{12})&= c_{\pm}^RF_R^{BW}e^{i\delta_{\pm}^R}P_l(\cos\theta),\\
\end{split}
\end{equation}
 where $F_R^{BW}$ is the the Breit-Wigner line-shape of the resonance $R$, 
 $c_{\pm}^R$ and $\delta_{\pm}^R$ are the corresponding
amplitude
and the relative phase, respectively.

By combining Eq. (\ref{Mtott1}) and the  relationships between  $\cos^n\theta$ and $P_l(\cos\theta)$, the  form of the  total amplitude squared can be derived as
\begin{equation} \label{Mtott4}
|\mathcal{M}_{\pm}^{\mathrm{tot}}|^2=\sum_{l=0}^{4}w_\pm^lP_l(\cos\theta),\\
\end{equation}
where the weights $w_{\pm}^l$ include both interference and non-interference contributions, and the projection weights for every component are  summarized in Table \ref{tabw}. With these, we can get
\begin{equation} \label{Mtott21}
\begin{split}
w_\pm^0&=\Bigg(|\mathcal{M}_{\pm}^{f_0}|^2+\frac{1}{5}|\mathcal{M}_{\pm}^{f_2}|^2+\frac{1}{3}|\mathcal{M}_{\pm}^{\rho^0}|^2\Bigg),\quad
w_\pm^1=\Bigg(2\mathrm{Re}[\mathcal{M}_{\pm}^{\rho^0}{\mathcal{M}_{\pm}^{f_0}}^*]+\frac{4}{5}\mathrm{Re}[\mathcal{M}_{\pm}^{\rho^0}{\mathcal{M}_{\pm}^{f_2}}^*]\Bigg)\\
w_\pm^2&=\Bigg(\frac{2}{3}|\mathcal{M}_{\pm}^{\rho^0}|^2+\frac{2}{7}|\mathcal{M}_{\pm}^{f_2}|^2+2\mathrm{Re}[\mathcal{M}_{\pm}^{f_0}{\mathcal{M}_{\pm}^{f_2}}^*]\Bigg),\quad w_\pm^3=\frac{6}{5}\mathrm{Re}[\mathcal{M}_{\pm}^{\rho^0}{\mathcal{M}_{\pm}^{f_2}}^*],\\
w_\pm^4&=\frac{18}{35}|\mathcal{M}_{\pm}^{f_2}|^2,\\
\end{split}
\end{equation}

\subsection{Results}\label{Results}
 
\begin{figure}[H]
\centering
\subfigure[]{
\includegraphics[width=.48\textwidth]{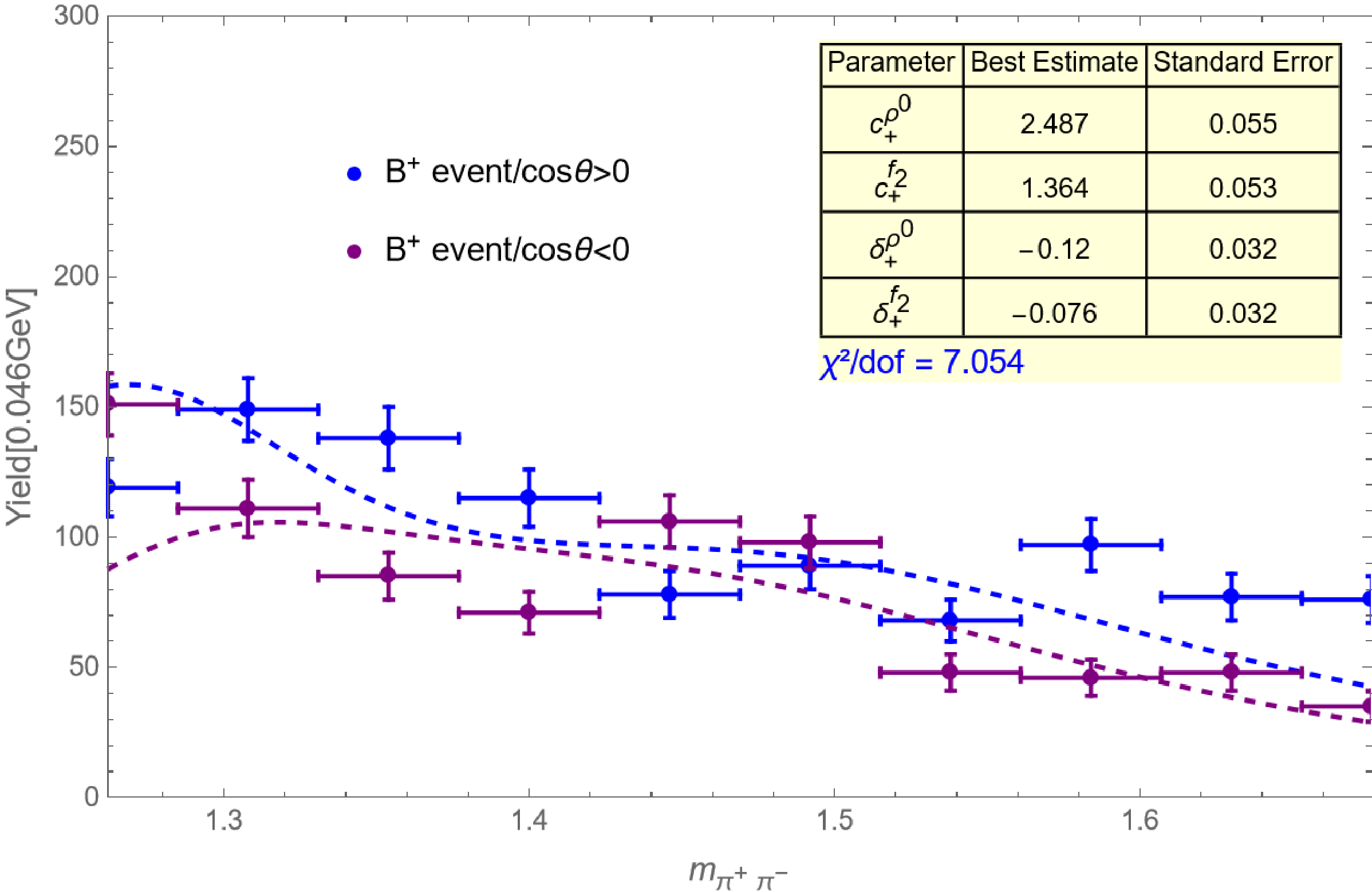}}
\subfigure[]{
\includegraphics[width=.48\textwidth]{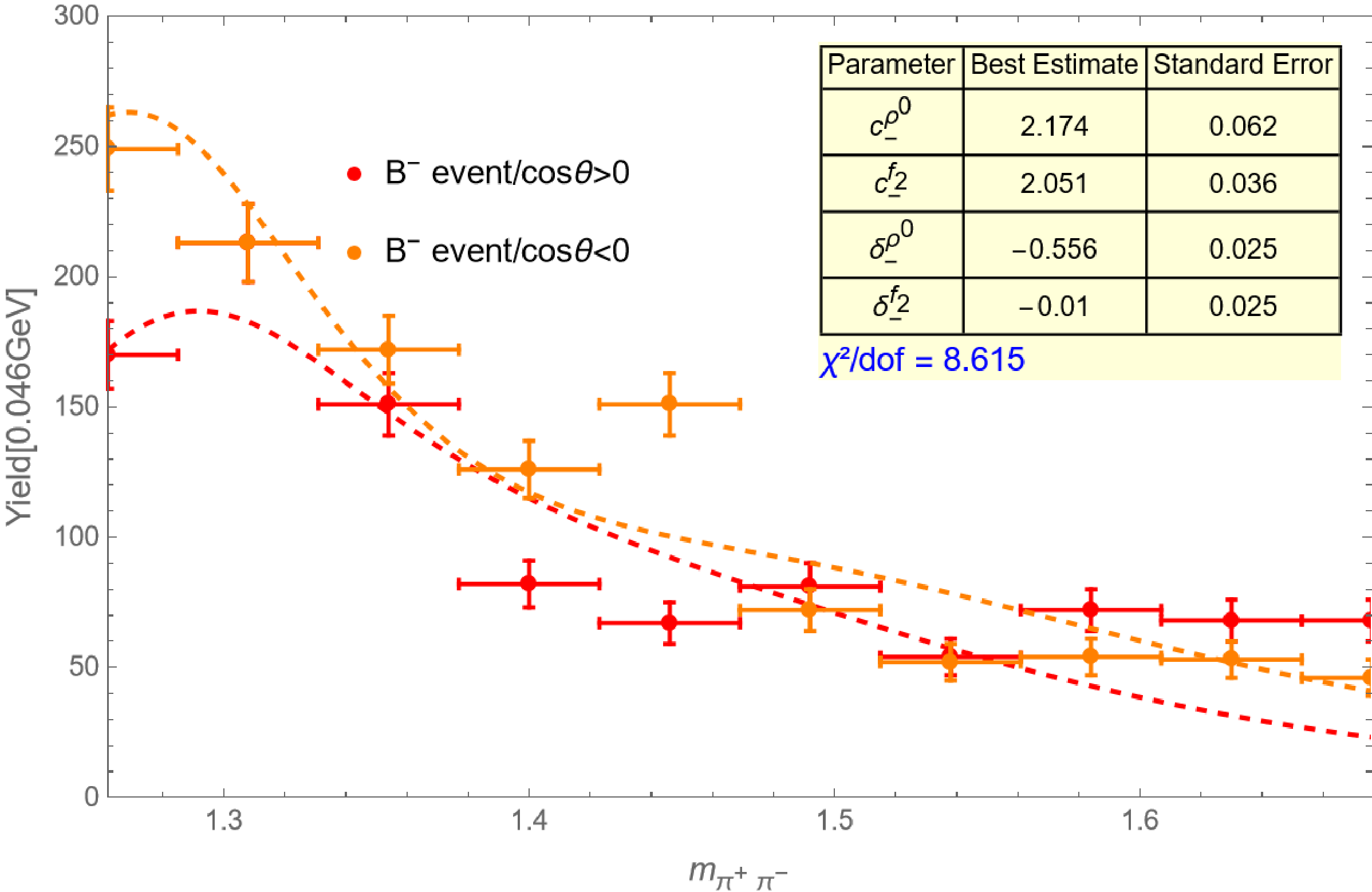}}
\caption{The  fitted results for  $B^+\rightarrow \pi^+\pi^+\pi^-$ (left) and $B^-\rightarrow \pi^-\pi^+\pi^-$ (right) decays in S1.}
\label{Events1}
\end{figure}

{\begin{table}[tb]
\renewcommand{\arraystretch}{1.2}
  \begin{center}
    \caption{The event yields for $B^\pm\rightarrow\pi^\pm\pi^+\pi^-$  with $m_{\pi\pi}$ ranging from 1.262 GeV to 1.676 GeV based on Fig. 12 in Ref. \cite{LHCb:2019sus}, denoted by $N_i(B^-)$ and $N_i(B^+)$ for $B^-\rightarrow\pi^- \pi^+\pi^-$ and $B^+\rightarrow\pi^+ \pi^+\pi^-$ decays, respectively, where the error propagation formula is adopted. Note that the helicity angle $\theta_{\mathrm{hel}}$ and $\theta$ mentioned in this paper satisfy the relation $\theta_{\mathrm{hel}}+\theta=180^\circ$.}

    \begin{tabular*}
    {\textwidth}
{@{\extracolsep{\fill}}c|cccccc}
    \hline\hline
    \multirow{2}{*}{Bin (GeV)} &\multicolumn{3}{c}{$\cos\theta_{\mathrm{hel}}\!\!>\!\!0$} & \multicolumn{3}{c}{$\cos\theta_{\mathrm{hel}}\!\!<\!\!0$}  \\
   \cline{2-4}
   \cline{5-7}
     & $N_i(B^-)-N_i(B^+)$ & $N_i(B^-)$  &  $N_i(B^+)$ & $N_i(B^-)-N_i(B^+)$ & $N_i(B^-)$  &  $N_i(B^+)$  \\
    \hline
    {1.262-1.285} &  $98\pm20$   &  $249\pm16$ & $151\pm12$  & $51\pm17$ &$170\pm13$  &$119\pm11$\\
        \hline
    {1.285-1.331} &  $102\pm18$      & $213\pm15$  &  $111\pm11$    & $64\pm19$   &  $213\pm15$ &$149\pm12$\\
        \hline
    {1.331-1.377} &  $87\pm16$     &  $172\pm13$  & $85\pm9$    & $13\pm17$   &  $151\pm12$& $138\pm12$ \\
        \hline
    {1.377-1.423} &  $55\pm14$     & $126\pm11$   & $71\pm8$    & $-33\pm14$   & $82\pm9$  & $115\pm11$\\
        \hline
    {1.423-1.469} &  $45\pm16$      & $151\pm12$   & $106\pm10$    & $-11\pm12$   & $67\pm8$ & $78\pm9$ \\
        \hline
    {1.469-1.515} &  $-26\pm13$      & $72\pm8$   & $98\pm10$    & $-8\pm13$   &  $81\pm9$ & $89\pm9$ \\
    \hline
    {1.515-1.561} &  $4\pm10$      & $52\pm7$   & $48\pm7$    & $-14\pm11$   &  $54\pm7$ & $68\pm8$ \\
    \hline
    {1.561-1.607} &  $8\pm10$      & $54\pm7$  & $46\pm7$    & $-25\pm13$   &  $72\pm8$ & $97\pm10$ \\
    \hline
    {1.607-1.653} &  $5\pm10$      & $53\pm7$  & $48\pm7$    & $-9\pm12$   &  $68\pm8$  &$77\pm9$ \\
\hline
    {1.653-1.676} &  $11\pm9$      & $46\pm7$   & $35\pm6$    & $-8\pm12$   &  $68\pm8$  &$76\pm9$ \\
 \hline\hline
\end{tabular*}\label{tab2}
  \end{center}
\end{table}

The parameter estimations are 
performed  by minimizing the $\chi^2$ 
function, using the binned event yields as the observables to be fitted. For S1, Eqs. (\ref{Mtott4})-(\ref{MRj}) and (\ref{events}) are adopted with $\cos{\theta}$ being restricted to either $[-1,0]$ 
or $[0,1]$. We set 
$c^{f_0}_\pm=0$ in Eq. (\ref{MRj}) to remove the  contributions from the $f_0(1370)$ or $f_0(1500)$ resonance. Theoretical results for the event yields in each bin are obtained by matching  the  data in in Ref. \cite{LHCb:2019sus} and are shown in Table \ref{tab2}.  The best-fit parameters $c_\pm^{\rho^0}$, $c_\pm^{f_2}$, $\delta_\pm^{\rho^0}$ and $\delta_\pm^{f_2}$,  together with the corresponding fitted yield distributions, are shown in Figs. \ref{Events1}(a) and \ref{Events1}(b) for the $B^+$ and $B^-$ decay channels,  respectively. For S2, we perform the fits with the $f_0(1370)$ and with $f_0(1500)$ resonances being added separately,  and the corresponding results are shown in Figs.
 \ref{EventsB1300} and \ref{EventsB1500}. Comparing Figs. \ref{EventsB1300} and \ref{EventsB1500}  suggests that including 
 $f_0(1500)$ may provide a better phenomenological description than including $f_0(1370)$. 

 \begin{figure}[H]
\centering
\subfigure[]{
\includegraphics[width=.48\textwidth]{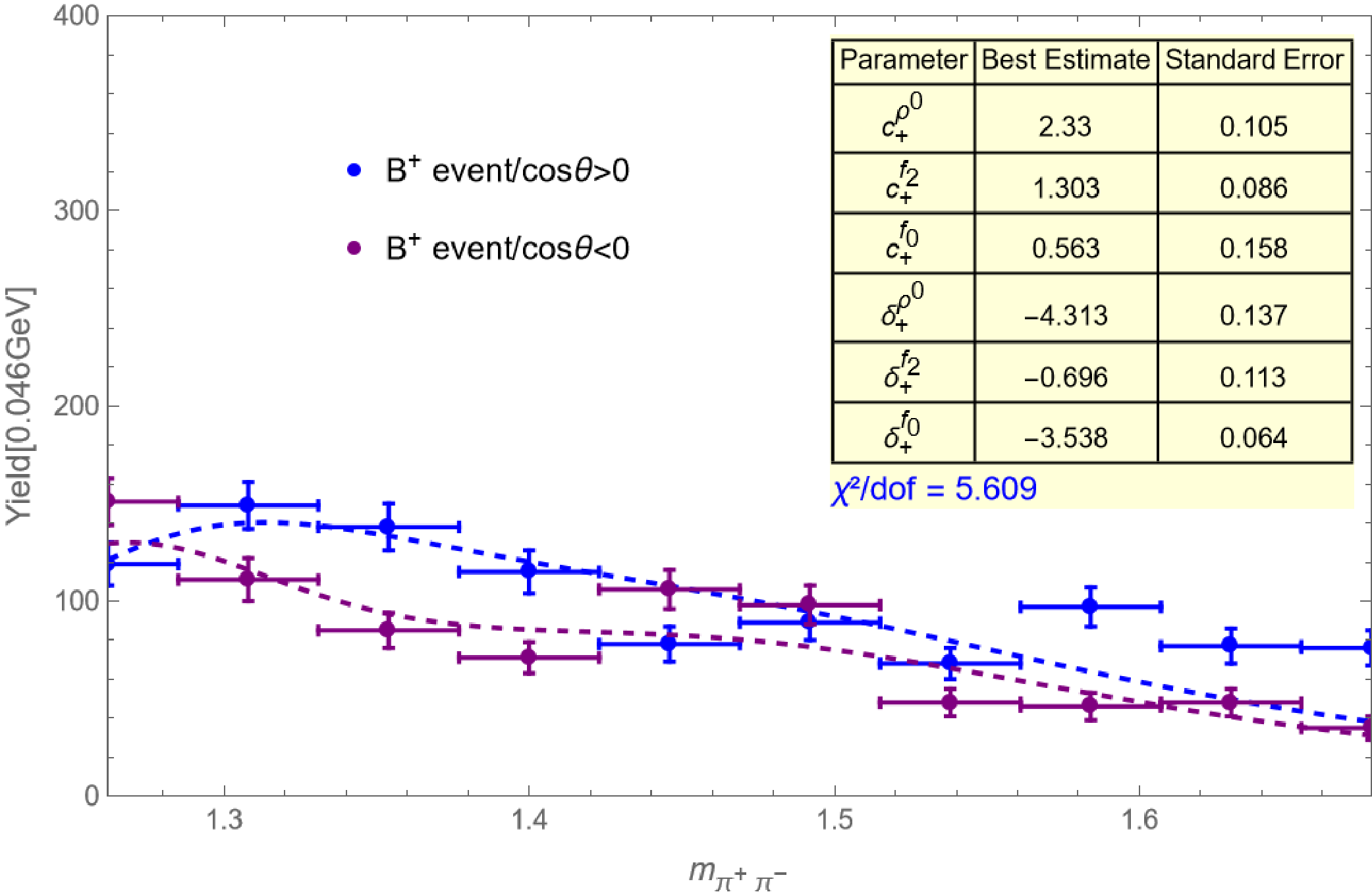}}
\subfigure[]{
\includegraphics[width=.48\textwidth]{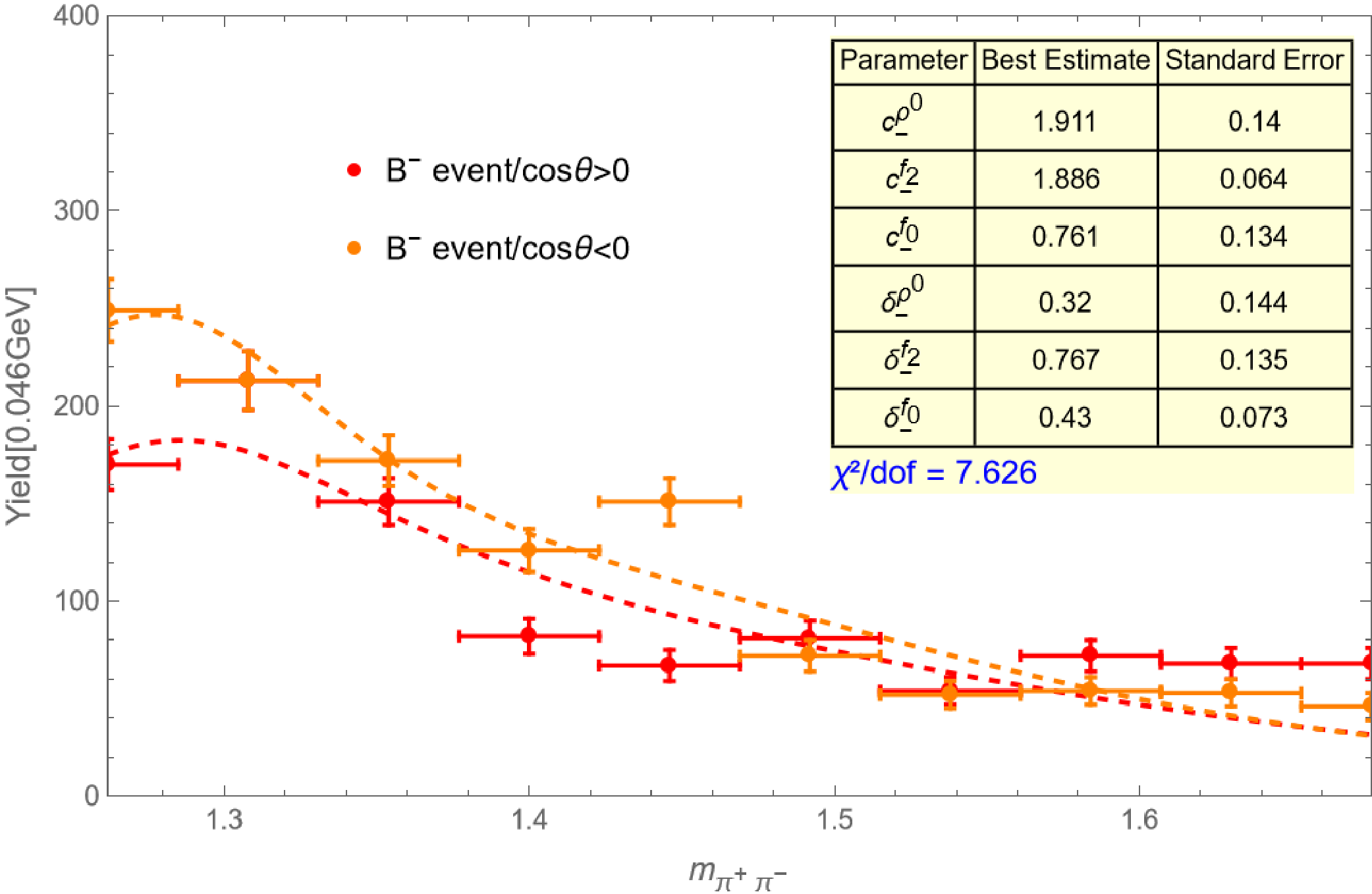}}
\caption{The same as  Fig. 1 except for adding the contribution from  the $f_0(1370)$ resonance in S2.}
\label{EventsB1300}
\end{figure}

\begin{figure}[H]
\centering
\subfigure[]{
\includegraphics[width=.48\textwidth]{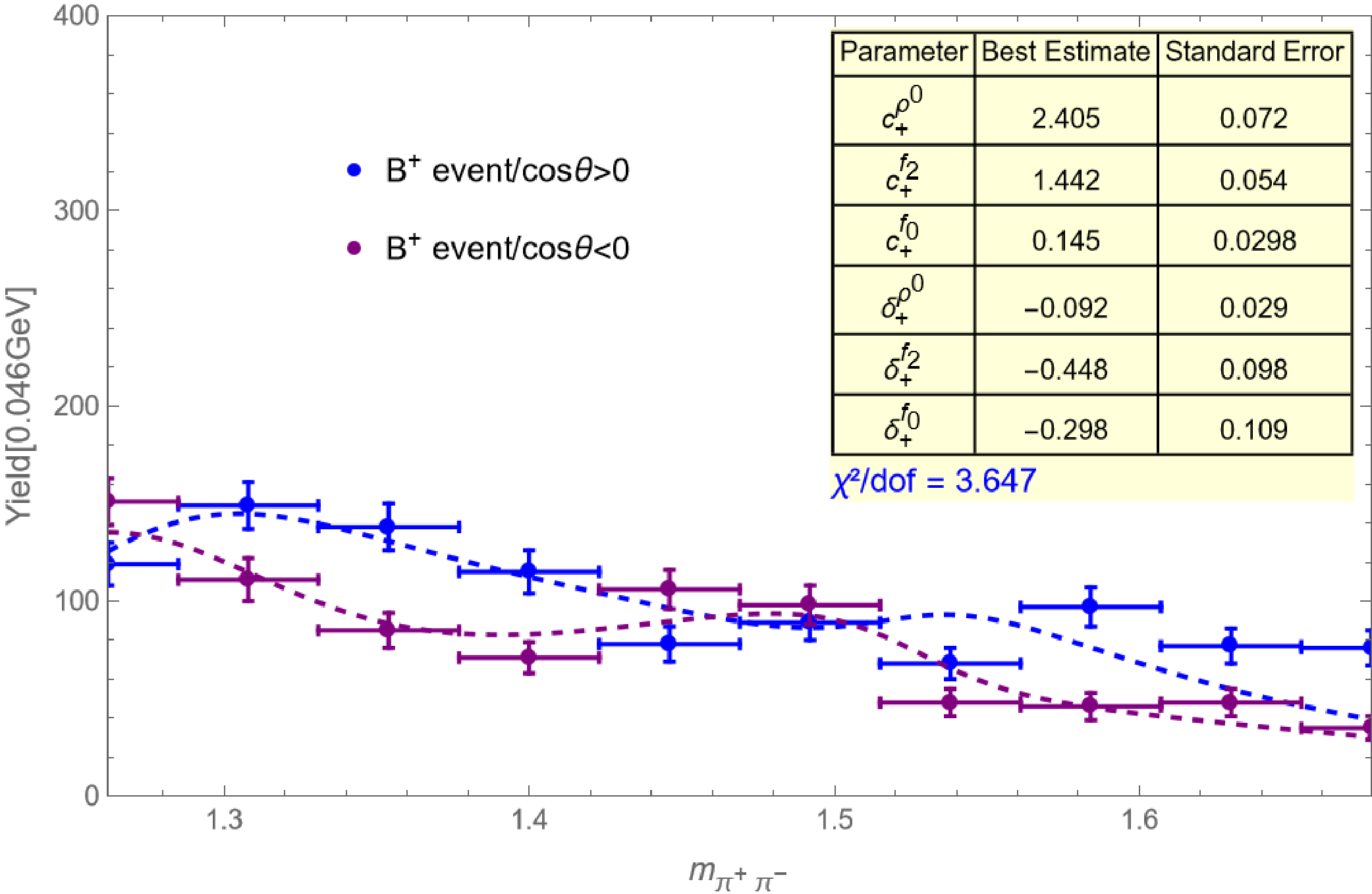}}
\subfigure[]{
\includegraphics[width=.48\textwidth]{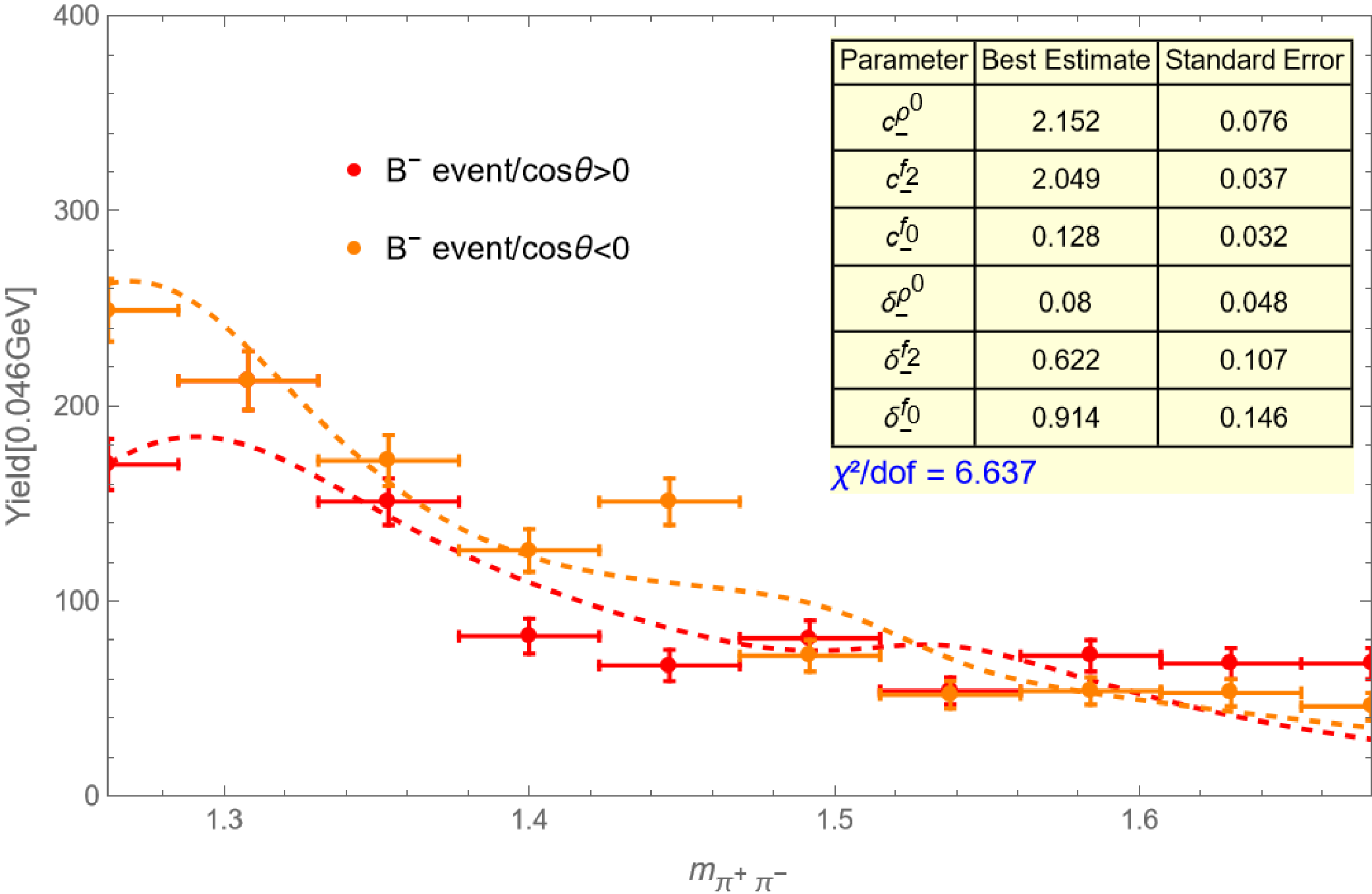}}
\caption{The same as Fig. 1 except for adding the contribution from the $f_0(1500)$ resonance in S2, which is also presented in Ref. \cite{Qi:2025zna}. }
\label{EventsB1500}
\end{figure}

\begin{figure}[H]
\centering
\subfigure[]{
\includegraphics[width=.48\textwidth]{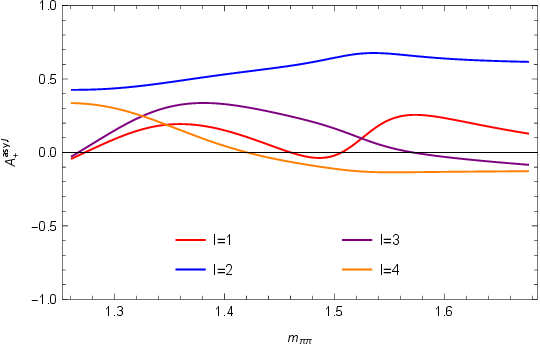}}
\subfigure[]{
\includegraphics[width=.48\textwidth]{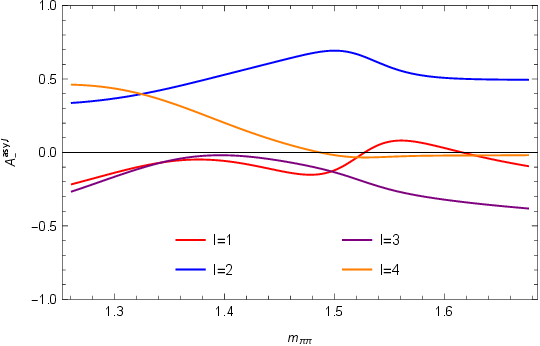}}
\caption{The results of  $\mathcal{A}_{\pm}^{\mathrm{asy},l}$ with adding the contribution from $f_0(1500)$ in S2.}
\label{ASyS2} 
\end{figure}

Substituting Eqs. 
(\ref{Mtott21}) and (\ref{Mtott4}) into Eq. (\ref{events}) shows explicitly that only the $l=0$ partial-wave expansion term   survives, whereas all higher components ($l = 1,2,3,4$) vanish. Consequently, both the 
non-interference contributions from $w_2$ and $w_4$, and  the interference terms from $w_1$, $w_2$ and $w_3$ in Eq. (\ref{Mtott21}) do not contribute to the event yields after the integration. To recover their contributions, we subdivide the phase space according to the zeros of the  Legendre polynomials and assign alternating signs to adjacent intervals by introducing the sgn(x) function in Eq. (\ref{eq:hatACPl}). For S2, using the fitted parameters in Fig.  \ref{EventsB1500} and Eqs. (\ref{Mtott1})-(\ref{Mtott21}), we evaluate Eq. (\ref{eq:hatACPl}) and obtain the $\mathcal{A}_{\pm}^{\mathrm{asy},l}$ observables. The results lie in the range about $[-40\%, 70\%]$ as shown in Fig. \ref{ASyS2} for $B^+$ and $B^+$ decays. As the $l$ varies, distinct resonance signals become clear, which are associated with $f_0(1500)$, $\rho^0(1450)$ and $f_2(1270)$, appear when
 $l = 2, 3,$ and $4$, respectively,
 while a signal from the $f_0(1500)$ state is also visible when $l = 1$.  The observables $\mathcal{A}_{\pm}^{\mathrm{asy},l}$ can be combined to obtain the corresponding CPA observables $\mathcal{A}_{\mathrm{CP}}^{\mathrm{asy},l}$,  as defined in Eq. (\ref{AasyCP}).  The theoretical results for $\mathcal{A}_{\mathrm{CP}}^{\mathrm{asy},l}$ are in the range about 
$[-20\%,9\%]$ as shown in Fig. \ref{S12ACP}. 
 
\begin{figure}[H]
\centering
\includegraphics[width=.48\textwidth]{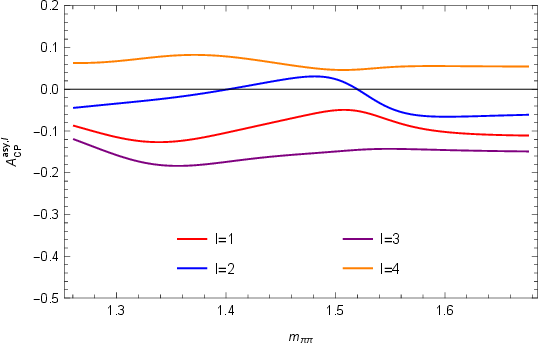}
\caption{The results of the $\mathcal{A}_{\mathrm{CP}}^{\mathrm{asy},l}$ obtained in S2 for the $B^\pm\rightarrow \pi^\pm\pi^+\pi^-$ decays.}
\label{S12ACP}
\end{figure}

To isolate the interference effects and compare them with the non-interference contributions in $\mathcal{A}_{\pm}^{\mathrm{asy},l}$ and $\mathcal{A}_{\mathrm{CP}}^{\mathrm{asy},l}$, we redefine two observables as $\mathcal{A}_{\pm}^{\mathrm{asy,int/non}}$ and $\mathcal{A}_{\mathrm{CP}}^{\mathrm{asy,int/non}}$, where the superscripts “ int” and “ non” denote contributions arising solely from the interference and non-interference terms, respectively, the index $l$ is hidden for brevity. To facilitate the comparison, the  quantities in Figs. \ref{ASyS2} and \ref{S12ACP} containing the summed contributions from both sources are labeled as $\mathcal{A}_{\pm}^{\mathrm{asy,tot}}$ and $\mathcal{A}_{\mathrm{CP}}^{\mathrm{asy,tot}}$. Specifically, each term in Eq. (\ref{Mtott21}) is separated into interference and non-interference components and then multiplied by the corresponding Legendre polynomial. Adopting Eqs.
(\ref{eq:hatACPl}) and  (\ref{AasyCP}), we compute $\mathcal{A}_{\pm}^{\mathrm{asy,int/non}}$ and $\mathcal{A}_{\mathrm{CP}}^{\mathrm{asy,int/non}}$, the results are shown in Figs. \ref{AsyP}-\ref{AsyCP}, including the  results of $\mathcal{A}_{\pm}^{\mathrm{asy,tot}}$ and $\mathcal{A}_{\mathrm{CP}}^{\mathrm{asy,tot}}$. It can be seen from Figs. \ref{AsyP} and \ref{AsyN} that for $l=1$ and $3$, the interference contributions dominate, while the non-interference contributions vanishe, for 
$l=2$ and 4, the opposite situation is observed. From our analysis, this behavior primarily reflects the adopted region-subdivision scheme, defined by the zeros of the Legendre polynomials, together with the parity properties of the polynomials.  For the odd-$l$, the interference and non-interference contributions are proportional to the
$P_{1,3}$ and $P_{0,2,4}$ terms, respectively. After the phase-space integration, the $P_{0,2,4}$
terms vanish. For even-$l$, the interference and non-interference parts are proportional to the
$P_{2}$ and  $P_{0,2,4}$ terms, respectively. In this case, the phase-space integrations of 
  $P_{0,2,4}$
are nonzero, so both contributions remain and their relative magnitudes depend on the specific theoretical predictions.  Consequently, the  choices of the
$l=1$ and $3$ are well suited to isolating interference contributions, whereas
$l=2$ and $4$, particularly 
$l=4$, are more suitable for studying the non-interference effects.

\begin{figure}[H]
\centering
\subfigure[]{
\includegraphics[width=.48\textwidth]{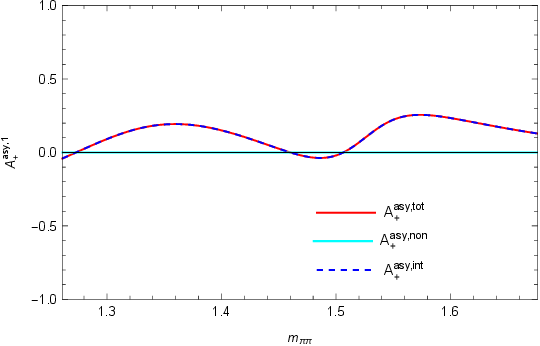}}
\subfigure[]{
\includegraphics[width=.48\textwidth]{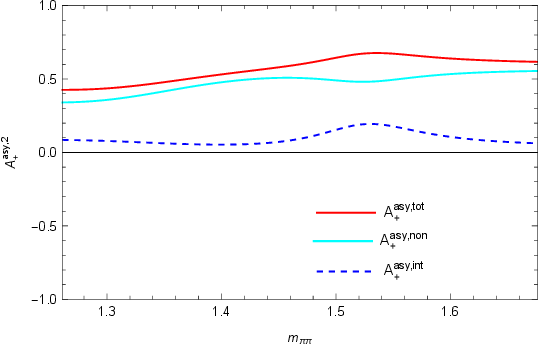}}
\subfigure[]{
\includegraphics[width=.48\textwidth]{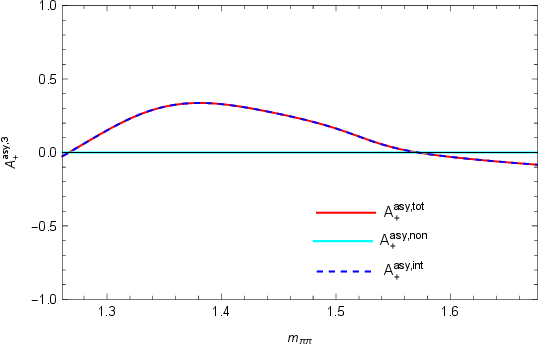}}
\subfigure[]{
\includegraphics[width=.48\textwidth]{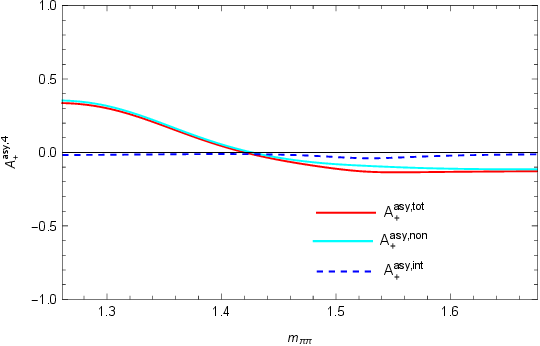}}
\caption{The results of  $\mathcal{A}_{+}^{\mathrm{asy,int}}$, $\mathcal{A}_{+}^{\mathrm{asy,non}}$ and $\mathcal{A}_{+}^{\mathrm{asy,tot}}$ with adding the contribution from $f_0(1500)$ in S2.}
\label{AsyP} 
\end{figure}

\begin{figure}[H]
\centering
\subfigure[]{
\includegraphics[width=.48\textwidth]{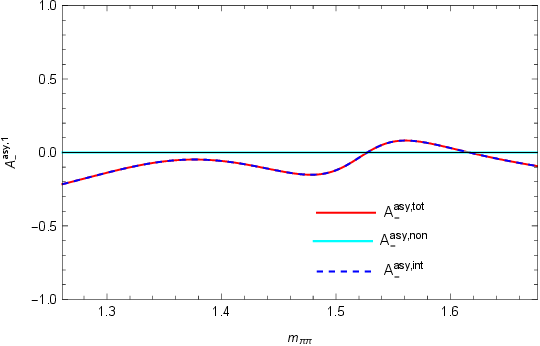}}
\subfigure[]{
\includegraphics[width=.48\textwidth]{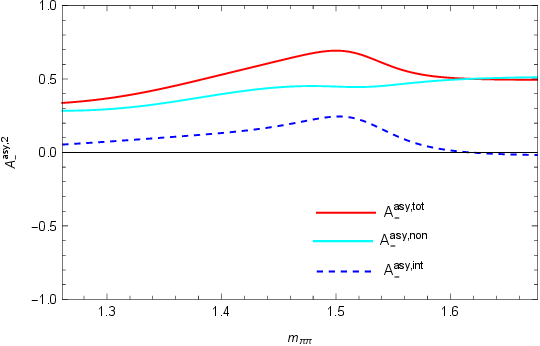}}
\subfigure[]{
\includegraphics[width=.48\textwidth]{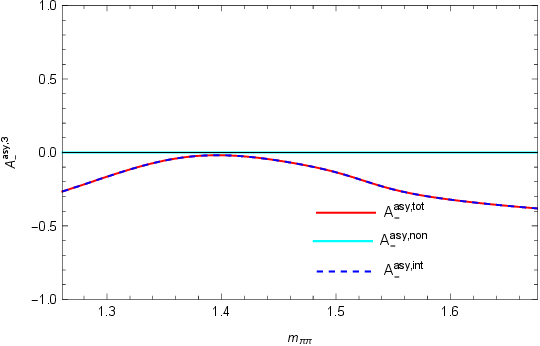}}
\subfigure[]{
\includegraphics[width=.48\textwidth]{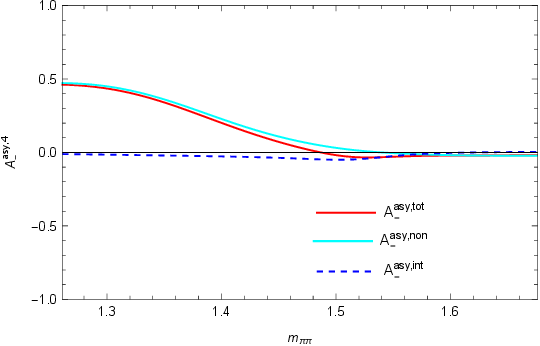}}
\caption{The results of  $\mathcal{A}_{-}^{\mathrm{asy,int}}$, $\mathcal{A}_{-}^{\mathrm{asy,non}}$ and $\mathcal{A}_{-}^{\mathrm{asy,tot}}$ with the contribution from $f_0(1500)$ in S2.}
\label{AsyN} 
\end{figure}

\begin{figure}[H]
\centering
\subfigure[]{
\includegraphics[width=.48\textwidth]{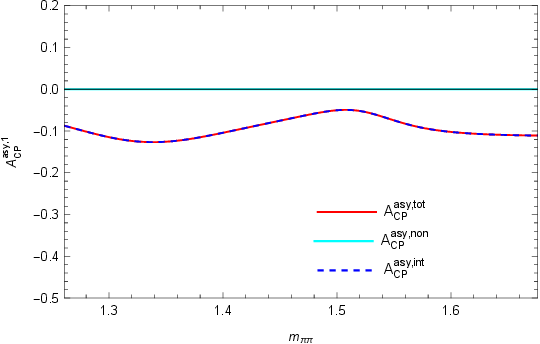}}
\subfigure[]{
\includegraphics[width=.48\textwidth]{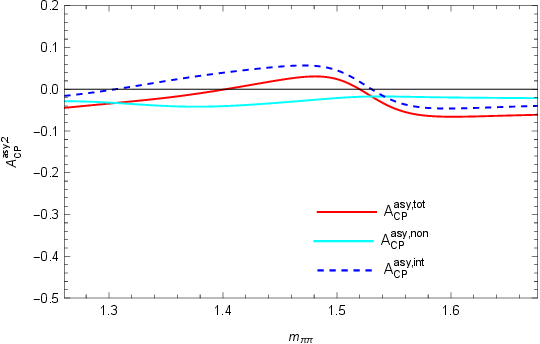}}
\subfigure[]{
\includegraphics[width=.48\textwidth]{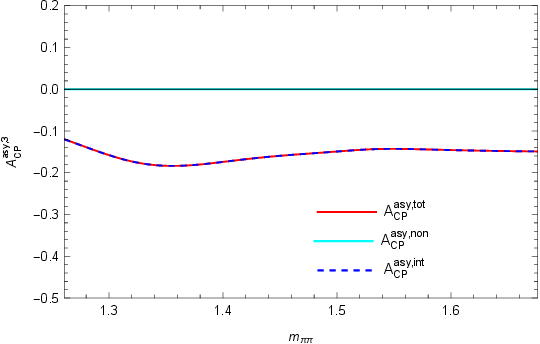}}
\subfigure[]{
\includegraphics[width=.48\textwidth]{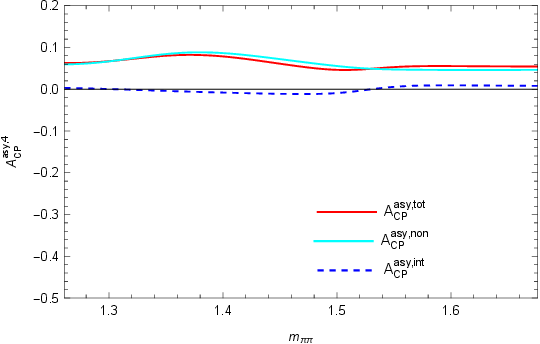}}
\caption{The results of  $\mathcal{A}_{CP}^{\mathrm{asy,int}}$, $\mathcal{A}_{CP}^{\mathrm{asy,non}}$ and $\mathcal{A}_{CP}^{\mathrm{asy,tot}}$ with  the contribution from $f_0(1500)$ in S2.}
\label{AsyCP} 
\end{figure}

\section{Conclusions}
CP violation in multi-body decays of heavy mesons contains rich dynamical information associated with both resonance structures and their interferences. In the conventional treatment, however, integrations over the full angular phase space remove the contributions from higher-order terms in the partial wave expansion of the total decay amplitude squared, thereby reducing or even eliminating the sensitivity to interference effects among different intermediate states. To address these limitations, within the partial wave analysis framework, we introduce a class of angular observables constructed with sign functions of Legendre polynomials, together with the corresponding CP asymmetries, denoted by 
$\mathcal{A}_{\pm}^{\mathrm{asy},l}$ and $\mathcal{A}_{\mathrm{CP}}^{\mathrm{asy},l}$, respectively. As an illustrative example, $B^\pm\rightarrow \pi^\pm \pi^+\pi^-$ decays are considered. Depending on whether the $\pi\pi$ $S$-wave resonances influence the study region, we adopt two research schemes and  the comparison shows that including 
$f_0(1500)$ provides a more satisfactory description. The theoretical results for 
$\mathcal{A}_{\pm}^{\mathrm{asy},l}$ and $\mathcal{A}_{\mathrm{CP}}^{\mathrm{asy},l}$ are obtained with the fitted parameters, shown in Figs. \ref{ASyS2} and \ref{S12ACP}. We further quantify the relative impacts of the interference and non-interference components with the aids of  $\mathcal{A}_{\pm}^{\mathrm{asy,int/non}}$ and $\mathcal{A}_{\mathrm{CP}}^{\mathrm{asy,int/non}}$, which results are presented in Figs. \ref{AsyP}–\ref{AsyCP}. Through comparison, it is found that for $l=1$ and 3, $\mathcal{A}_{\pm}^{\mathrm{asy,l}}$ and $\mathcal{A}_{\mathrm{CP}}^{\mathrm{asy,l}}$ are dominated by interference contributions, whereas the situation is opposite for $l=2$ and 4. Such behavior originates from the angular structure of the contributions. 
The interference (non-interference) contributions are proportional to 
$P_{1,3}$ and ($P_{0,2,4}$) for odd 
$l$ and to 
$P_{2}$ ($P_{0,2,4}$) for even 
$l$. Since the $P_{0,2,4}$ terms vanish under phase-space integration only in the odd-
$l$ case, odd-
$l$ observables isolate the interference contribution, whereas even-
$l$ ones retain significant sensitivity to non-interference terms. Their relative magnitudes depend on specific theoretical predictions.These
findings can be expected to generalize to other heavy-hadron decay processes.

\section*{Acknowledgments}
This work was supported by  National Natural Science Foundation of China (12405115, 12105149, 12475096, 12275024 and the High-Level Scientific Research Fund of Ningbo University under Grants No. GJPY2026032).

\clearpage
\bibliography{ref.bib}
\end{document}